\documentclass[aps,amsmath,showpacs,amsfonts,10pt]{revtex4}
\usepackage{epsfig,graphicx}
\usepackage[english]{babel}
\usepackage{amsfonts}
\usepackage{amsmath}
\usepackage{latexsym}
\usepackage{graphics,bm}
\usepackage{subfigure}
\usepackage{dcolumn}
\usepackage{bm}
\usepackage{rotating}

\begin{document}

\title{Transport behaviour of a Bose Einstein condensate in a bichromatic optical lattice}

\author{Aranya B Bhattacherjee$^{*}$ and Monika Pietrzyk$\dag$}

\address{$^{*}$Max Planck-Institute f\"ur Physik komplexer Systeme, N\"othnitzer Str.38, 01187 Dresden,Germany}

\address{$\dag$Weierstra\ss-Institut f\"ur Angewandte Analysis und Stochastik, Mohrenstr.39, 10117 Berlin, Germany}

\begin{abstract}
The Bloch and dipole oscillations of a Bose Einstein condensate (BEC) in an optical
superlattice is investigated. We show that the effective mass increases in an optical
superlattice, which leads to localization of the BEC, in accordance with recent experimental
observations [16]. In addition, we find that the secondary optical lattice is a useful
additional tool to manipulate the dynamics of the atoms.
\end{abstract}


\maketitle

\section{Introduction}

Interference pattern of intersecting laser beams create a periodic potential for atoms, which
is known as an optical lattice \cite{Morsch06}. Ultracold bosons trapped in such periodic potentials have
been widely used recently as a model system for the study of some fundamental concepts of
quantum physics. Josephson effects \cite{Anderson98}, squeezed states \cite{Orzel01}, Landau-Zener tunneling and Bloch
oscillations \cite{Morsch01} and superfluid-Mott insulator transition \cite{Greiner02} are some examples. An important
promising application under study is quantum computation in optical lattices \cite{Sebby06}. Optical
lattices are therefore, of particular interest from the perspective of both fundamental quantum
physics and its connection to applications.
Using superposition of optical lattices with different periods \cite{Peil03}, it is now possible to
generate periodic potentials characterized by a richer spatial modulation, the so-called
optical superlattices. The light-shifted potential of the superlattice is described as
\begin{equation}
V(z)=V_{1}\cos^{2}\left(\frac{\pi z}{d_{1}} \right)
+V_{2}\cos^{2}\left(\frac{\pi z}{d_{2}}+\phi \right).
\end{equation}
Here $d_{1}$ and $d_{2}$ are, respectively, the primary and secondary lattice constants,
$V_{1}$ and $V_{2}$ are the respective amplitudes and $\phi$ is the phase of the secondary lattice.
When $\phi=0$, all sites of the lattice are perfectly equivalent due to the symmetries of the
system, so that the population and onsite energies are same at each site. An asymmetry is
introduced when $\phi\not=0$ and hence the onsite energies are not the same at each site.

Theoretical interest in optical superlattice started only recently. These include work on
fractional filling Mott insulator domains \cite{Buonsante04}, dark \cite{Louis04} and gap \cite{Louis05} solitons, the Mott-Peierls
transition \cite{Dimt03}, non-mean field effects \cite{Rey04} and phase diagram of BEC in two color superlattices
\cite{Roth03}. In a recent work, the analogue of the optical branch in solid-state physics was predicted
in an optical superlattice \cite{Huang05}. Rousseau et al. \cite{Marcos06} have considered the effect of a secondary lattice on an one dimensional hard core bosons (strongly correlated regime). A detailed theoretical study of the Bloch and Bogoliubov
spectrum of a BEC in a one-dimensional optical superlattice has been done \cite{Bhat07}. In a very
recent experiment \cite{Lye06}, it was observed that the center of mass motion of a BEC is
blocked in a quasi-periodic lattice. Considering the fact that these optical superlattices are
now being realized experimentally and interesting experiments are being done routinely, we were
motivated to study the influence of the secondary lattice on Bloch oscillations and dipole
oscillations of atoms.

\section{Bloch Oscillations}

We consider an elongated cigar shaped BEC confined in a harmonic trap potential of the form $V_{ho}(r,z)=\frac{m}{2}(\omega^{2}_{r}r^{2}+\omega^{2}_{z}z^{2})$
and a one-dimensional tilted optical superlattice of the form $V_{op}(z)=E_{R}\left( s_{1}\cos^{2} (\frac{\pi z}{d})
+s_{2} \cos^{2} (\frac{\pi z}{2d})\right)+mgz $. We have taken a particular case of $d_{2}=2 d_{1} =2 d$. Here $s_{1}$ and $s_{2}$ are the dimensionless amplitudes of the primary and the secondary superlattice potentials with $s_{1}>s_{2}$. $E_{R}=\dfrac{\hbar^{2} \pi^{2}}{2 m d^{2}}$ is the recoil energy ( $\omega_{R}=\dfrac{E_{R}}{\hbar} $ is the corresponding recoil frequency) of the primary
lattice. We take $\omega_{r}>>\omega_{z}$ so that an elongate cigar shaped BEC is formed.
The harmonic oscillator frequency corresponding to small motion about the minima of the
optical superlattice is $\omega_{s}\approx \dfrac{\sqrt{s_{1}}\hbar \pi^{2}}{m d^{2}}$.
The peak densities in each well match the Gaussian profile. Since the array is tilted, the
atoms undergo coherent Bloch oscillations driven by the interwell gravitational potential $mgz$.
The BEC is initially loaded into the primary lattice and the secondary lattice is switched
on slowly. The frequency of each minima of the primary lattice is not perturbed significantly
by the addition of the secondary lattice. Here $\omega_{s}>>\omega_{z}$ so that the optical lattice
dominates over the harmonic potential along the $z$-direction and hence the harmonic potential
is neglected. The strong laser intensity will give rise to an array of several
quasi-two-dimensional pancake shaped condensates. Because of the quantum tunneling,
the overlap between the wave functions of two consecutive layers can be sufficient to
ensure full coherence. We study now the Bloch dynamics of the BEC in the tilted optical
superlattice by solving the discrete nonlinear schroedinger equation (DNLSE). The dynamics
of the BEC is governed by the Gross-Pitaevskii equation (GPE),
\begin{equation}
i\hbar \dfrac{\partial \zeta}{\partial t}=-\dfrac{\hbar^{2}}{2m} \nabla^{2} \zeta
+\left\lbrace V_{ho}(r,z)+V_{op}(z)+g_{0}|\zeta|^{2} \right\rbrace \zeta,
\label{GPE}
\end{equation}
where $g_{0}=\dfrac{4\pi \hbar^{2} a}{m}$, with $a$ the two body scattering length and $m$
the atomic mass. In the tight binding approximation the condensate order parameter can be
written as
\begin{equation}
\zeta(r,t) = \sqrt{N_T} \sum_j \Psi_j(t) \phi(r-r_j),
\label{order}
\end{equation}
where $N_T$ is the total number of atoms and $\phi(r-r_j) = \phi_j$ is the condensate
wavefunction localized in the trap $j$ with $\int dr \phi_j \phi_{j+1} \approx 0$, and
$\int dr \left| \phi_j \right|^2 = 1$; $\Psi_j(t)$ is the $j^{th}$ amplitude.
$\Psi_j(t) = \sqrt{\rho_j(t)} \exp(i \theta_j(t))$ where $\rho_j=\frac{N_j}{N_T}$, with
$N_j$ and $\theta_j$ being the number of particles and phases in the trap $j$ respectively.
Substituting the Ansatz (\ref{order}) in (\ref{GPE}), we find that the GPE reduces to the DNLSE, 
\begin{equation}
i \frac{\partial \Psi_j}{\partial t} =
- \frac12 \left \{ \left(1-\alpha (-1)^{j-1} \right) \Psi_{j-1}
+ \left(1- \alpha(-1)^{j} \right) \Psi_{j+1} \right \}
+ \left(\varepsilon_j + \Lambda \left| \Psi_j\right|^2 \right) \Psi_j.
\label{DNLSE}
\end{equation}
Here $\varepsilon_j = \frac{1}{J_0}
\int dr \left[ \frac{\hbar^2}{2m} \left(\bar{\nabla} \phi_j \right)^2
+ \left(V_{ho}(r) + V_{op}(z)\right) \left|\phi_j\right|^2 \right]$,
$\Lambda= \frac{g_0 N_T}{J_0} \int dr \left|\phi_j\right|^4$,
$\alpha=\frac{\Delta_0}{2 J_0}$. One can show using
$J_j = - \int dr \left[ \frac{\hbar^2}{2m} \bar{\nabla} \phi_j \cdot \bar{\nabla} \phi_{j+1}
+ \phi_j \left(V_{ho}(r) + V_{op}(z) \right) \phi_{j+1} \right]$ that there are distinctly
two Josephson coupling parameters, $J_{1,2}=J_0 \pm \frac{\Delta_0}{2}$ where
$J_0 \approx \frac{E_R}{4} \left[ \left(\frac{\pi^2}{2} -2 \right) s_1 \right]
\exp\left(- \frac{\pi^2 \sqrt{s_1}}{4} \right)$ and
$\Delta_0 \approx \frac{E_R}{2} s_2 \exp\left(- \frac{\pi^2 \sqrt{s_1}}{4} \right)$
\cite{Bhat07}.
We have rescaled time as $t \rightarrow \frac{\hbar}{2 J_0} t$. In Eq. (\ref{DNLSE}),
$\varepsilon_j = \omega_B j$, where $\omega_B = \frac{mg\lambda_1}{4 J_0}$ is the frequency
of Bloch oscillation and $\lambda_1$ is the wavelength of the laser creating the primary
lattice. In order to understand the Bloch and dipole oscillations, we solve the DNLSE using
a variational approach adopted from \cite{Trom01}. The Hamiltonian function corresponding to the DNLSE Eq. (\ref{DNLSE}) reads
\begin{equation}
\hspace*{-1mm} H = \sum_j \left[ \frac{-1} {4} \left \{ \left(1-(-1)^j \alpha\right)
\left(\Psi_j\Psi_{j+1}^* + \Psi_j^* \Psi_{j+1} \right)
+ \left(1-(-1)^{j-1} \alpha \right)
\left(\Psi_j\Psi_{j-1}^* + \Psi_j^* \Psi_{j-1} \right) \right \}
+ \varepsilon_j \left|\Psi_j \right|^2 + \frac{\Lambda}{2} \left|\psi_j \right|^4 \right],
\end{equation}
where $\sum_j \left| \Psi_j \right|^2 = 1$. To analyze the Bloch dynamics, we study the
dynamical evolution of a site dependent Gaussian wavepacket, which we parameterize as
\begin{equation}
\Psi_j(t) = \sqrt{K} \exp\left[ -\frac{(j-\xi)^2}{\gamma^2} + ip \left(j-\xi \right)
+ i \frac{\delta}{2} \left(j-\xi \right)^2 + i (-1)^j \frac{\phi}{2} \right],
\end{equation}
where $\xi(t)$ and $\gamma(t)$ are, respectively, the center and width of the condensate,
$p(t)$ and $\delta(t)$ are their associated momenta, and $K(\gamma, \xi)$ a normalization factor.
Here $(-1)^j \frac{\phi}{2}$ is the phase of the wave packet at the $j^{th}$ site. Clearly,
depending upon whether $j$ is odd or even, the phase is $\pm \frac{\phi}{2}$. As explained in
ref.(15), as the condensate moves from one well to the next, it acquires additional phase,
which depends on the height of the barrier. As the height of the barrier alternates,
the phase also alternates.

The dynamics of the wave packet can be obtained by the variational principle from the Lagrangian,
$L = \sum_j i \Psi_j \Psi_j^* -H$, with the equations of motion for the variational parameters
$q_i(t) = \xi, \gamma, p, \delta, \phi$ given by
$\frac{d}{dt} \frac{\partial L}{\partial \dot q_i} = \frac{\partial L}{\partial q_i}$.
The phase is used to enforce a constraint. The Lagrangian is derived as
\begin{equation}
L= p \dot{\xi} - \frac{\gamma^2 \dot{\delta}}{8}
- \left[\frac{\Lambda}{2 \sqrt{\pi}\gamma}\right]
+\left \{ \cos{\phi} \cos{p} + \alpha \sin{\phi} \sin{p} \right \}
\exp(- \eta) - V(\gamma, \xi),
\end{equation}
where $\eta= \frac{1}{2 \gamma^2} + \frac{\gamma^2 \delta^2}{8}$ and
$V(\gamma, \xi) = K \int_{-\infty}^{\infty} dj  \, \varepsilon_j
\exp\left(-2 \frac{(j-\xi)^2}{\gamma^2} \right)$. 

The variational equations of motion are derived as:

\begin{subequations}
\begin{equation}
\dot p = -\frac{\partial V}{\partial \xi},
\label{variatA}
\end{equation}
\vspace*{-7mm}
\begin{equation}
\dot{\xi} = \left[\cos{\phi} \sin{p} - \alpha \sin{\phi} \cos{p} \right] \exp(-\eta),
\label{variatB}
\end{equation}
\vspace*{-7mm}
\begin{equation}
\dot{\delta} = \left[\cos{\phi} \cos{p} + \alpha \sin{\phi} \sin{p} \right] \exp(-\eta)
\left[\frac{4}{\gamma^4} - \delta^2 \right] + \frac{2 \Lambda}{\gamma^3 \sqrt{\pi}}
- \frac{4}{\gamma} \frac{\partial V}{\partial \gamma},
\label{variatC}
\end{equation}
\vspace*{-7mm}
\begin{equation}
\dot{\gamma} = \gamma \delta
\left[\cos{\phi} \cos{p} + \alpha \sin{\phi} \sin{p} \right] \exp(-\eta),
\label{variatD}
\end{equation}
\vspace*{-7mm}
\begin{equation}
\tan{\phi} = \alpha \tan{p}.
\label{variatE}
\end{equation}
\end{subequations}
Since $\cos^2{\phi} + \sin^2{\phi} = 1$, together with equation (\ref{variatA}-\ref{variatE}),
we get the following constraints on $\cos{\phi}$ and $\sin{\phi}$:
\begin{subequations}
\begin{equation}
\cos{\phi} = \frac{\cos{p}}{\sqrt{\cos^2{p} + \alpha^2 \sin^2{p}}},
\label{constrainsA}
\end{equation}
\vspace*{-7mm}
\begin{equation}
\cos{\phi} = \frac{\alpha \sin{p}}{\sqrt{\cos^2{p} + \alpha^2 \sin^2{p}}}.
\label{constrainsB}
\end{equation}
\end{subequations}
Corresponding to the variational equations (\ref{variatA}-\ref{variatE}) and constraints (\ref{constrainsA}-\ref{constrainsB})
the effective Hamiltonian is written as
\begin{equation}
H= \frac{\Lambda}{2 \sqrt{\pi} \gamma} - \cos{p} \sqrt{1+\alpha^2 \tan^2 p}\exp(-\eta)
+V(\gamma,\xi).
\end{equation}
We first study the Bloch oscillations. For the tilted
periodic potential the on-site energies are written as $\varepsilon_j = j \omega_B$.

\begin{figure}[t]
\hspace{-2.0cm}
\includegraphics{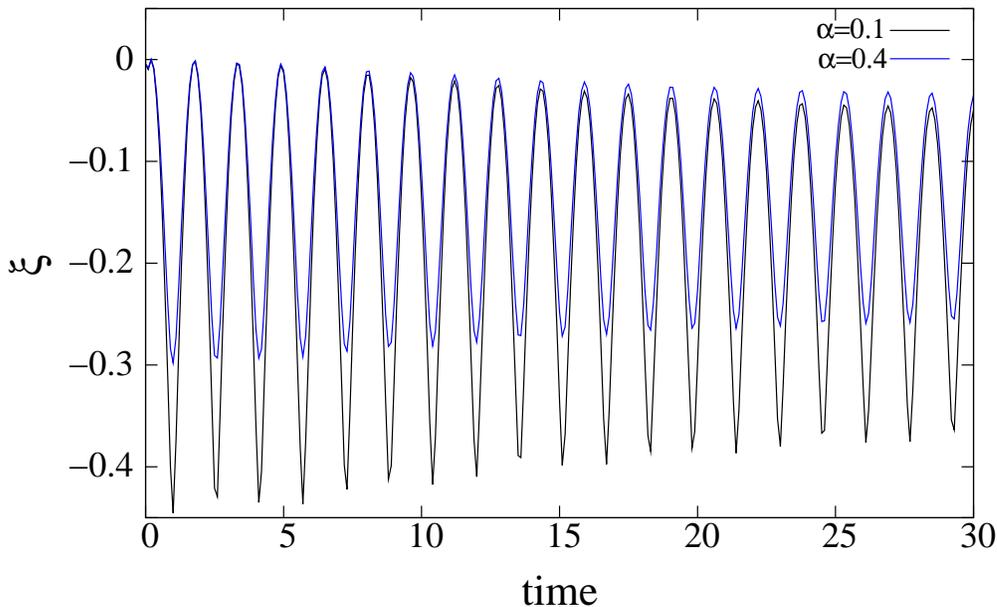} 
\caption{Oscillations of the center of mass $\xi(t)$ is depicted for two different values of the secondary lattice strength, $\alpha=0.1$ and $\alpha = 0.4$. The other parameters are $\xi(0) =0, \, p(0) = 0, \, \delta(0) =0, \, \gamma(0) = 10$, 
$\Lambda=20, \, \omega_B=2$. On increasing the strength of the secondary lattice, the amplitude of the center of mass motion reduces.}
\label{fig:figure_1}
\end{figure}

Using equations (\ref{variatA}-\ref{variatE}), we find $V = \xi \omega_B$ and $\dot{p} = - \omega_B$.
We solve the variational equations of motion numerically for the following initial values
$\xi(0) =0, \, p(0) = 0, \, \delta(0) =0, \, \gamma(0) = 10$ and the parameters
$\Lambda=20, \, \omega_B=2$. The result for the center of mass $\xi(t)$ is depicted in
figure 1 for two different values of the secondary lattice strength, $\alpha=0.1$ and $\alpha = 0.4$. Clearly on increasing the strength of the secondary lattice from $\alpha=0.1$ to  $\alpha = 0.4$, the amplitude of the center of mass motion reduces. The secondary lattice serves to break the discrete translational invariance of the system, thus favouring localization of the wave function. Optical superlattices with higher periodicities will block the center of mass more strongly.
The observed damping (with respect to time) in fig.1 is due to interactions. In the absence
of interactions, the center of the BEC for $p_0 = 0$ goes roughly as
$\xi(t) \approx - (1-\alpha^2) (1-\cos{\omega_B t})$, while in the presence of
interactions, the oscillations roughly decreases as
$\xi(t) \approx - (1-\alpha^2) \left(1-\exp\left(-\frac{\Lambda t^2}{2 \pi \gamma_f^4} \right)
\cos{\omega_B} t \right)$. Here, $\gamma_f$ is some final value of $\gamma$. Clearly when
there is no interaction, there is no damping of the Bloch oscillations in time but 
there is a reduction in the amplitude by a factor $(1-\alpha^2)$ due to the presence of the
secondary lattice. In order to understand the origin of this blocking of the center of mass
motion, we derive the effective mass $(m^*)^{-1} = \frac{\partial^2 H}{\partial p^2}$ as,
\begin{equation}
m^* = \frac{\left(1+ \alpha^2 \tan^2{p}\right)^{3/2} \exp(\eta)}
{\cos{p} \left(1-\alpha^2 \tan^4{p} \right) (1-\alpha^2)}.
\end{equation}
A diverging effective mass $m^* \rightarrow \infty$ as $t \rightarrow \infty$ due to
interactions leads to a self-trapping of the wave packet [17]. In the expression for the
effective mass (eqn.11), in the absence of interaction, the factor $\exp(\eta)$
is constant since $\gamma$ tends to a final value $\gamma_f$
and $\delta(t) \approx \delta_0$ (initial value). This can be seen from equations 8c and 8d.
The effective mass is now enhanced due to the presence of the secondary lattice. Since
$\Lambda = 0$, the effective mass stays constant in time and the Bloch oscillations show
reduced oscillations compared to the case for a single frequency optical lattice but does not
show damping in time. On the other hand when $\Lambda \neq 0$, and $t \rightarrow \infty$,
$\gamma \rightarrow \gamma_f$ and $\delta(t) \approx \frac{2 \Lambda t}{\gamma^3_f \sqrt{\pi}}$,
so that $m^* \rightarrow \infty$. This causes not only a reduction in amplitude but also
damping in time. It is interesting to note that, we now have an additional handle to tune
the effective mass. A plot between $m^*$ and $\alpha$ (for $p=0$) in fig.2 shows that as the
strength of the secondary lattice increases, the effective mass also increases. Therefore the
origin of the reduction of the amplitude of Bloch oscillations of a BEC in an optical superlattice
is due to an increase of the effective mass. Dynamics of localized excitations, such as solitons depends on the effective mass, hence the secondary lattice emerges as a useful additional handle to manipulate localized excitations.

\begin{figure}[t]
\hspace{-2.0cm}
\includegraphics{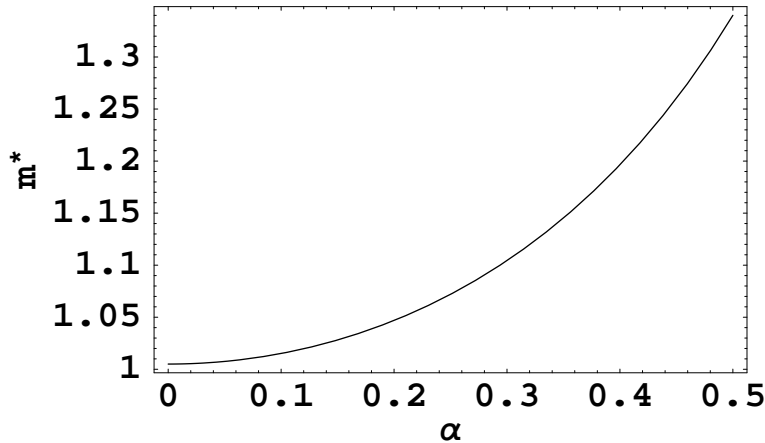} 
\caption{A plot between $m^{*}$ and $\alpha$ shows that as the strength of the secondary lattice increases, the effective mass also increases. Therefore the origin of the localization of a BEC in an optical superlattice is due to an increase in the effective mass.}
\label{fig:figure_2}
\end{figure}

\section{Dipole Oscillations}

We study now the dipole oscillations. Instead of the gravitational potential, we consider
a sufficiently large ($\omega_z \approx \omega_s$) magnetic harmonic potential superimposed
on the optical lattice, $\varepsilon_j = \Omega j^2$, where $\Omega = \frac{m \omega^2_z d^2}{J_0}$. The variational equations of motion give $V(\gamma, \xi) = \Omega \left(\frac{\gamma^2}{4} + \xi^2\right)$ and $\dot{p} = -2 \Omega \xi$.
In the regime of negligible mean field interaction ($\Lambda = 0$) and small momenta $p$,
the equation for the center of mass is $\dot{\xi}(t) = \left(1-\alpha^2 \right)p $.
Consequently, the center of mass obeys the equation of an undamped harmonic oscillator,
$\ddot{\xi} = \omega_d^2 \xi$, where the frequency of dipole oscillation,
$\omega_d^2 = 2 \Omega \left(1-\alpha^2 \right) = \omega_z^2 \left(\frac{m}{m^*} \right)$
is reduced in the presence of the secondary lattice since $m^{*}>m$. We consider the initial conditions
$\xi(0) = 0$ and $p(0) = p_0$. The center of mass in the $\Lambda =0$
regime and small momenta is
$\xi(t) \approx \frac{\left(1 - \alpha^2 \right)^{1/2}}{\sqrt{2 \Omega}} \sin{\omega_d} t$.
In the low momenta limit, the amplitude of the center of mass decreases with increasing
strength of the secondary lattice approximately as
$\left[ 1-\left(\frac{s_2}{s_1 \left( \frac{\pi^2}{2} -2 \right)} \right)^2 \right]^{1/4}$. In the experiment of ref.[16],
$\omega_z \approx 2 \pi \times 10$ Hz and $\lambda_1 \approx 830 \times 10^{-9}$ nm.
This corresponds to a very low value of $\Omega$ $\approx 0.0001$ (in dimensionless units).

\begin{figure}[t]
\hspace{-2.0cm}
\includegraphics{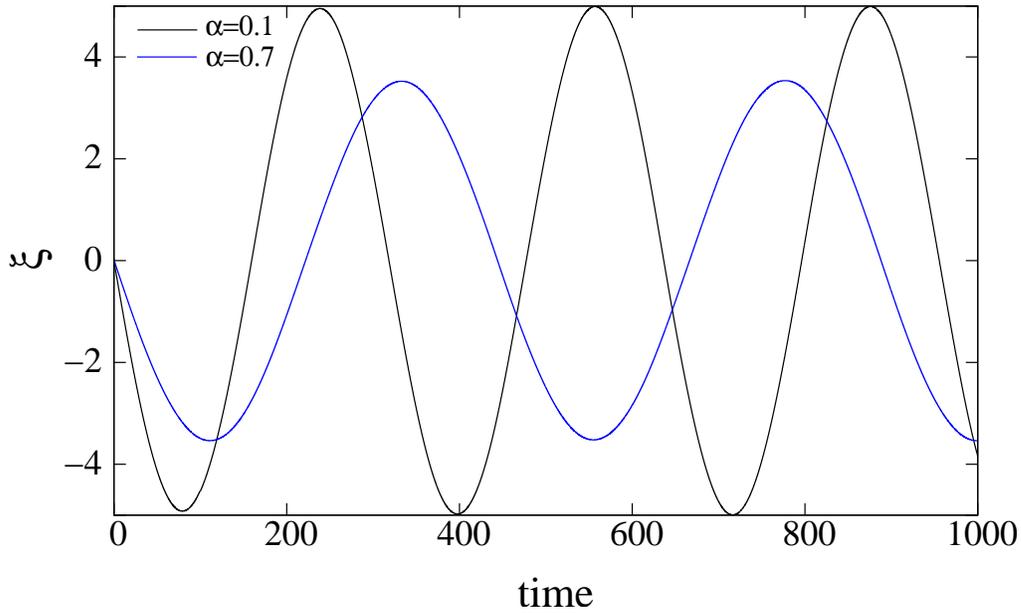} 
\caption{A plot of the dipole oscillations for $\alpha=0.1$ and $\alpha = 0.7$. The other parameters are $\xi(0) =0, \, p(0) = 0.1, \, \delta(0) =0, \, \gamma(0) = 40$, $\Lambda=5, \, \Omega=.0002$. We notice that increasing the strength of the secondary lattice, the dipole oscillations are blocked, in accordance with the experimental observations of \cite{Lye06}. Since, we are in the negligible mean field interaction regime, the dipole oscillations are not damped. }
\label{fig:figure_3}
\end{figure}

We solve the variational equations of motion numerically for the following initial values:
$\xi(0) =0, \, p(0) = 0.1, \, \delta(0) =0, \, \gamma(0) = 40$ and the parameters:
$\Lambda=5, \, \Omega=.0002$. The result for the dipole oscillation is depicted in
figure 3 for two different values of the secondary lattice strength $\alpha=0.1$ and $\alpha = 0.7$. For $\Lambda=5$, we are still in the regime of negligible mean field interaction and we do not expect any damping. On increasing the strength of the secondary lattice, the amplitude of the center of mass $\xi(t)$ is reduced in accordance with the experiments of \cite{Lye06}. This reduction in the amplitude of the dipole oscillation on increasing the strength of the secondary lattice is due to an increase in the
effective mass, as mentioned earlier in this paper. The initial value of the effective mass can be positive ($\cos{p_0} > 0$)
or negative ($\cos{p_0} <0 $). Let us suppose that $\cos{p_0} >0$
and initial values: $\gamma(0)= \gamma_0$, $\delta(0)= \delta_0 = 0$ and $\xi(0)= \xi_0 =0$. The initial value of the
Hamiltonian is $H_0 = \frac{\Lambda}{2 \sqrt{\pi} \gamma_0} - \cos{p_0} \sqrt{1+ \alpha^2 \tan^2{p_0}}
\exp\left(- \frac 12 \gamma_0^2 \right) + \frac{\Omega \gamma_0^2}{4}$. Since the Hamiltonian
is conserved, we have $H_0 = \frac{\Lambda}{2 \sqrt{\pi} \gamma_0} - \cos{p_0} \sqrt{1+ \alpha^2 \tan^2{p_0}}
\exp\left(- \frac 12 \gamma^2 - \frac{\gamma^2 \delta^2}{8} \right)+\frac{\Omega \gamma^2}{4}$.
The parabolic external potential helps to keep $H_0 > 0$, therefore,
\begin{equation}
\frac{\Lambda}{2 \sqrt{\pi} \gamma} + \frac{\Omega \gamma^2}{4} - H_0 >0.
\end{equation}

The trajectories in the $\gamma -\delta$ plane are given by

\begin{equation}
\delta^2 = - \left[\frac{8 \gamma^2 \log \left(\frac{\frac{\Lambda}{2 \pi \gamma}
+ \frac{\Omega \gamma^2}{4} - H_0}{\cos{p_0} \sqrt{1+\alpha^2 \tan^2{p_0}}} \right) +4}
{\gamma^4} \right].
\end{equation}

\begin{figure}[t]
\hspace{-2.0cm}
\includegraphics{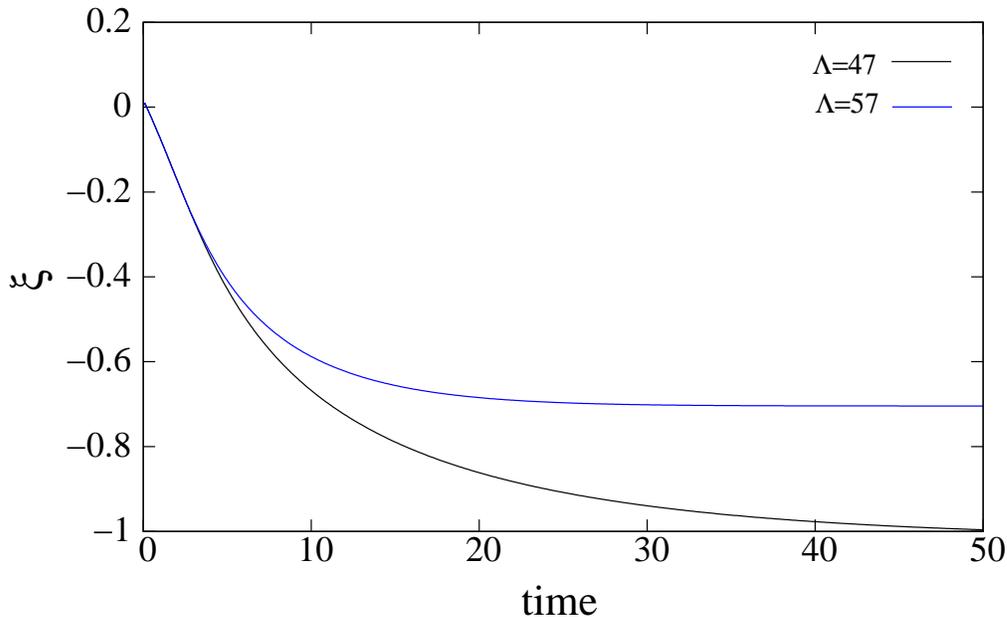} 
\caption{Center of mass motion for $\alpha = 0.1$ , $\xi(0)=0, \, p_0 = 0.1, \,\delta_0=0.1, \, \gamma_0 = 10, \, \Lambda =47,57, \, \omega=0.0002$. We notice that for such high values of $\Lambda$, the dipole oscillations are completely blocked. Both interactions and secondary lattice induced disorder cooperate to block the center of mass motion.}
\label{fig:figure_4}
\end{figure}

Fig. 4 shows a plot of the center of mass for $\alpha = 0.1$ , $\xi(0)=0, \, p_0 = 0.1, \,\delta_0=0.1, \, \gamma_0 = 10, \, \Lambda =47,57, \, \omega=0.0002$. We notice that for such high values of $\Lambda$, the dipole oscillations are completely blocked. Both interactions and secondary lattice induced disorder cooperate to block the center of mass motion. For the higher value of $\Lambda$,  the center of mass stops at an earlier time, which again is in accordance with experiments \cite{Lye06}. From equation (13), we notice that
$\delta \rightarrow \infty$ as $t \rightarrow \infty$. Therefore, for large time,

\begin{equation}
\dot{\xi} \approx (1-\alpha^2) \sin{p_0} \exp\left(- \frac{1}{2 \gamma_{max}^2}
- \frac{\gamma_{max}^2 \delta^2}{8} \right) \rightarrow 0
\end{equation}
and
\begin{equation}
m^* = \frac{\left( 1+ \alpha^2 \tan^2{p} \right)^{3/2}
\exp\left(\frac{1}{2 \gamma_{max}^2} + \frac{\gamma_{max}^2 \delta^2}{8} \right)}
{\cos{p} \left(1 - \alpha^2 \tan^4{p} \right) (1-\alpha^2)} \rightarrow \infty.
\end{equation}

The center of the BEC wavepacket stops and the effective mass goes to
infinity and there is an energy transfer from the kinetic energy to the internal modes,
since $\delta$ is the momentum associated with the width $\gamma$. This is the self trapped
regime. We also find that the final value of center of mass $\xi_f$ is not the same as $\xi_0$. For a fixed $\Lambda$, an increase in the secondary lattice potential will block the center of mass at an earlier time.

\section{Conclusions}

In conclusion, we have studied the Bloch and dipole oscillations of a Bose Einstein condensate trapped in an optical superlattice. In particular, we have shown that due to the addition of the secondary lattice, the center of mass motion is blocked which leads to a blockage of the center of mass motion. This effect is due to an increase in the effective mass in the presence of the secondary lattice. The frequency of the dipole oscillations is also found to be reduced due to the secondary lattice. These results are in accordance with recent experiments \cite{Lye06}. The secondary lattice is found to be a promising tool to investigate and manipulate localized excitations.

\begin{acknowledgments}
One of the authors A.B acknowledges support from the German Academic Exchange Service (DAAD) for the fellowship (A/06/33410) and is grateful to Professor R. Graham, for providing the facilities for carrying out part of this work at the University of Duisburg-Essen, Germany. The author A.B is also grateful to the Max Planck Institute for Physics of Complex Systems, Dresden, Germany for the hospitality. M.P acknowledges the support by the project D20 in the DFG Research Center {\sc Matheon} "Mathematics for key technologies" in Berlin.
\end{acknowledgments}

\end{document}